\documentclass[twocolumn,aps,prx,superscriptaddress,floatfix,notitlepage]{revtex4-1}
\usepackage[latin9]{inputenc}
\usepackage{color}
\usepackage{textcomp}
\usepackage{amsmath}
\usepackage{amssymb}
\usepackage{graphicx}
\usepackage{esint}

\makeatletter
\usepackage{amsfonts}
\usepackage{bbold}
\usepackage{natbib}

\pdfoutput=1
\newcommand{\ket}[1]{|#1 \rangle}
\newcommand{\bra}[1]{\langle#1 |}

\newcommand{\lr}[1]{\left( #1 \right)}

\newcommand{\mean}[1]{\langle #1 \rangle}
\newcommand{\no}{\nonumber}
\newcommand{\tr}[1]{\mathrm{tr}\left\{#1\right\}}

\begin{document}

\title{Photonic Quantum Circuits with Time Delays}

\author{Hannes Pichler}

\affiliation{Institute for Quantum Optics and Quantum Information of the Austrian
Academy of Sciences, A-6020 Innsbruck, Austria}

\affiliation{Institute for Theoretical Physics, University of Innsbruck, A-6020
Innsbruck, Austria}

\author{Peter Zoller}

\affiliation{Institute for Quantum Optics and Quantum Information of the Austrian
Academy of Sciences, A-6020 Innsbruck, Austria}

\affiliation{Institute for Theoretical Physics, University of Innsbruck, A-6020
Innsbruck, Austria}
\begin{abstract}We study the dynamics of photonic quantum circuits consisting of nodes coupled by quantum channels. We are interested in the regime where time delay in communication between the nodes is significant. This includes the problem of quantum feedback, where a quantum signal is fed back on a system with a time delay. We develop a matrix product state approach to solve the Quantum Stochastic Schrödinger Equation with time delays, which accounts in an efficient way for the entanglement of nodes with the stream of emitted photons in the waveguide, and thus the non-Markovian character of the dynamics. We illustrate this approach with two paradigmatic quantum optical examples: two coherently driven distant atoms coupled to a photonic waveguide with a time delay, and a driven atom coupled to its own output field with a time delay as an instance of a quantum feedback problem.
\end{abstract}

\date{\today}
\maketitle

\begin{figure}
\includegraphics[width=0.5\textwidth]{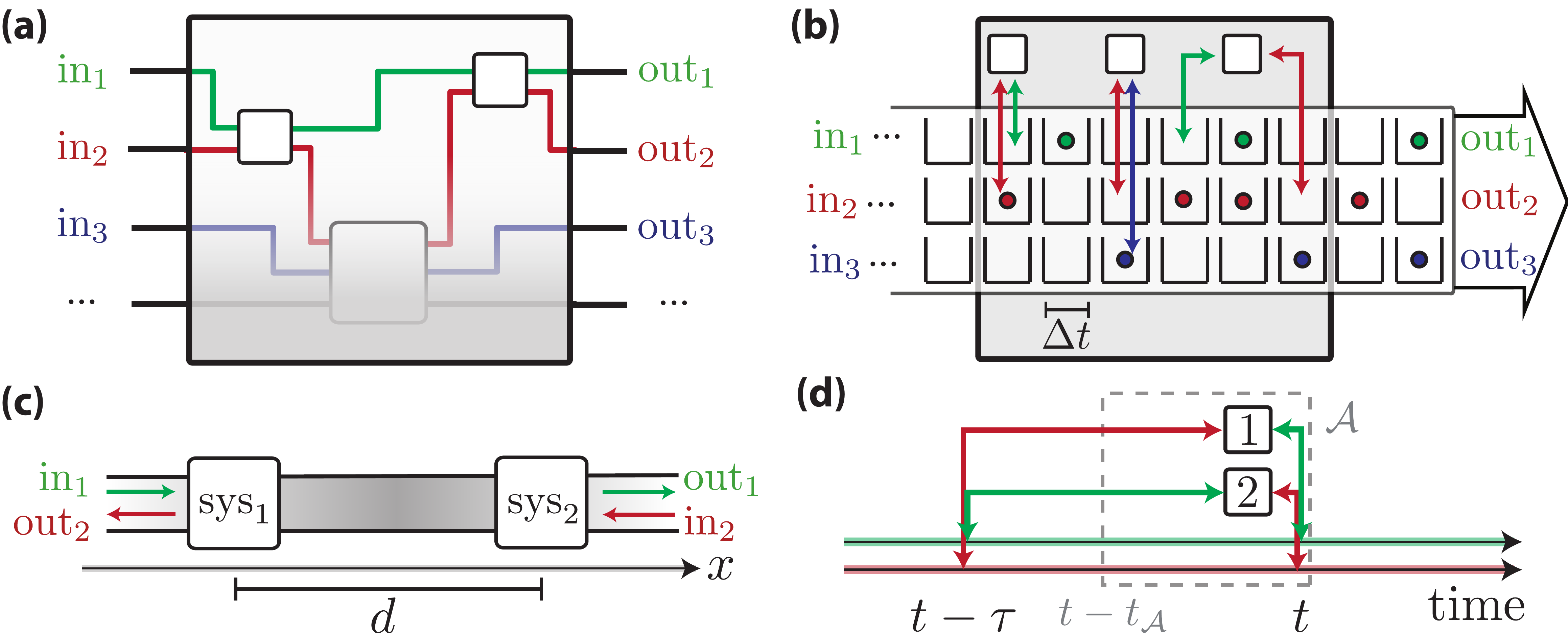}\caption{(a) Photonic quantum circuit consisting of waveguides (quantum channels) connecting the nodes (boxes), and representing the input and output ports of the system.  While (a) is a {\em spatial representation} of the circuit, (b) is the corresponding {\em time interpretation} according to a QSSE  \eqref{eq:QSSEOp1}. The nodes interact sequentially with photon modes defined for {\em time bins} according to the stroboscopic map \eqref{eq:QSSEmap}. The time delay $\tau>0$ corresponds to a {\em non-local} interaction in time. (c) Simple network consisting of 
two systems (atoms) at distance $d$ coupled to a 1D wave guide, and (d) the time interpretation according to QSSE \eqref{Eq2-1}. The red/green line represent the left/right moving radiation field in the time representation (see text).}
\label{Fig1} 
\end{figure}

\textit{Introduction.} Wiring up increasingly complex quantum devices
from basic modules is central in the effort to build large scale quantum
circuits  \cite{Kimble:2008if}. Quantum optical systems provide a natural framework to implement such a modular approach as a photonic quantum circuit \cite{Hucul:2014fm,Schoelkopf:2008cs}. Here left and right propagating modes in optical fibers or waveguides provide the channels for communication between the nodes, and represent input
and output ports to drive and observe the circuit {[}Fig.~\ref{Fig1}(a){]}.
Such networks can involve quantum communication between 'local' nodes,
or in a distributed network between 'distant' nodes, where time delays
can be important. Recent advances in building small scale quantum
processors with atoms and ions \cite{Duan:2010dq}, and the development of atom-photon
interfaces in CQED \cite{Tiecke:vw}, or in coupling atoms to photonic nanostructures \cite{Mitsch:2014fz,Sollner:2015fc,Goban:2014eq},
have demonstrated - at least on a conceptual level - the basic building
blocks for such a scalable photonic network \cite{Santori:2014bk,RamosVermersch}.

On the theory side this
raises the question of formulating a quantum theory of photonic quantum
networks. Such a theory must account for the quantum many-body dynamics
induced by multiple photon exchanges between the nodes, and relating
the input and output quantum signals on the level of quantum states.
Theoretical quantum optics has provided tools for modeling {\em
Markovian} quantum networks, i.e.~when time delays can be ignored \cite{QWI,QWII}.  
It is the purpose of the present work to address non-Markovian aspects of the dynamics introduced by these time delays. 
This refers to both retarded interaction between the nodes of the network involving the exchange of (possibly many) photons, and also addresses the problem of {\em quantum feedback} \footnote{We contrast this to feedback where a measurement is performed and we act back on the quantum system \cite{Wiseman:2010vw}}, where the quantum signal emitted from a  system is fed back with a time delay \footnote{Here we are interested in time delays in contrast to the non-Markovianity from structured reservoirs \cite{Rivas:2014bl,BreuerReview,Strunz:1999ew}.}.  Our approach is based on solving the Quantum Stochastic Schrödinger Equation (QSSE) \cite{QWII} with time-delays based on (continuous) matrix
product states (MPS) \cite{Schon:2005fk,Schon:2007bw,Osborne:2010hr,Verstraete:2010bf,Haegeman:2010bc}, as developed originally in a condensed matter context \cite{Fannes:1992vq,White:1992zz,Ostlund:1995fx,Vidal:2004jc,Daley:2004hk,Schollwock:2005wz,Peropadre:2013ia}. This technique allows for an efficient description of entanglement, which scales with finite time delays between the nodes of the network and the stream of photons propagating in the quantum channels, including the quantum fields and relevant observables at the output of the photonic quantum network. 

\textit{Quantum optical model}. Our approach is best illustrated for
the paradigmatic model \cite{Milonni:1974bo} consisting of two distant nodes $n=1,2$ connected to an infinite
waveguide, representing the input and output ports of our system (c.f.~Fig.\ref{Fig1}(c)).
The nodes are located at positions $x_{1}<x_{2}$, and there is a
time delay $\tau=(x_{2}-x_{1})/v\equiv d/v$ associated with photon
exchange with $v$ the velocity of light. Our treatment considers
photons in a bandwidth ${\cal B}$ around some mean optical frequency
$\bar{\omega}$. To describe the dynamics, we follow the familiar
quantum optical formulation \cite{QWII,SUPP} and
write a QSSE $i\hbar\frac{d}{dt}\ket{\Psi}=H(t)\ket{\Psi}$ for the
total system of nodes and waveguide. The total Hamiltonian is denoted
by $H(t)=H_{{\rm sys}}+{H}_{{\rm int}}(t)$ with $H_{{\rm sys}}=\sum_{n}H_{{\rm sys}}^{(n)}$
the sum of the Hamiltonian of the nodes. The interaction term is given
by
\begin{align}
{H}_{{\rm int}}(t) & \!=\!i\hbar\!\left(\!\big(\sqrt{\gamma_{L}}b_{L}^{\dag}\!\lr{t}\!+\!\sqrt{\gamma_{R}}b_{R}^{\dag}\!\lr{t-\tau}\!e^{i\phi}\big)c_{1}\!-\!\textrm{h.c.}\!\right)\no\label{Eq2-1}\\
& +\!i\hbar\!\left(\!\big(\sqrt{\gamma_{L}}b_{L}^{\dag}\!\lr{t-\tau}\!e^{i\phi}\!+\!\sqrt{\gamma_{R}}b_{R}^{\dag}\!\lr{t}\!\big)c_{2}\!-\!\textrm{h.c.}\!\right)
\end{align}
which describes the emission and absorption of photons by the nodes
$n=1,2$ into the left and right-propagating modes $i=L,R$ of the
waveguide within a rotating wave approximation (RWA). Here the operators $b_{R}(t)$ and $b_{L}(t)$ are defined
by 
\[
b_{i}(t)=\frac{1}{\sqrt{2\pi}}\int_{\mathcal{B}}d\omega\,b_{i}(\omega)e^{-i(\omega-\bar{\omega})t}\quad(i=L,R)
\]
with $b_{i}(\omega)$ ($b_{i}^{\dagger}(\omega)$) destruction (creation)
operators of photons of frequency $\omega$, satisfying $[b_{i}(\omega),b_{i'}^{\dag}(\omega')]=\delta_{i,i'}\delta(\omega-\omega')$.
In quantum optics they have the meaning of quantum noise operators
with white noise bosonic commutation relations $[b_{i}(t),b_{i'}^{\dag}(t')]=\delta_{i,i'}\delta(t-t')$, which lead to the interpretation as a QSSE. The photon propagation phase shift is denoted by $\phi=-\bar{\omega}\tau$. The operators $c_{1}$and
$c_{2}$ are the transition operators for the node in emission of
a photon. The coupling to the left and right propagating modes is given
by the decay rates $\gamma_{L}$ and $\gamma_{R}$ into the radiation
modes of the waveguide, respectively, and we define $\gamma\equiv \gamma_R+\gamma_L$. We note that for a {\em  chiral} coupling, as is naturally realized in nanophotonics \cite{Mitsch:2014fz,Sollner:2015fc,Goban:2014eq},
we have $\gamma_{R}\ne\gamma_{L}$.
While the assumptions behind the derivation of the QSSE are still Markovian in nature \cite{QuantumNoiseBook,QWII}, it is the time delays,  reflecting retardation between the absorption and emission events, which introduces a non-Markovian element into the dynamics [see Fig.~\ref{Fig1}(d)].

The simplest physical realization of two nodes ($n=1,2$) is given by coherently driven two level systems with ground and excited states $\ket{g_n}$, $\ket{e_n}$ \cite{Mitsch:2014fz}. 
The corresponding system Hamiltonian is in a rotating frame $H_{\rm sys}^{(n)}=-\hbar\Delta_n\ket{e_n}\bra{e_n}-\frac{\hbar}{2}(\Omega_n \ket{g_n}\bra{e_n}+{\rm h.c.})$, with $\Delta_n=\omega_L-\omega_{eg}$ the detuning of the driving laser from atomic resonance, $\Omega_n$ the Rabi frequency, and $c_n\equiv \sigma_n^{-} = \ket{g_n}\bra{e_n}$. We return to this example below in some detail. 

In the Markovian limit, $\tau\!\rightarrow\!0^{+}$,
the above QSSE can be interpreted according to Stratonovich quantum
stochastic calculus \cite{QuantumNoiseBook,QWII}. There are well established techniques to convert
this QSSE to  Ito calculus, and eventually to a master equations
for the dynamics of the reduced state of the nodes, $\rho_{\rm sys}$, tracing over the waveguide as a quantum reservoir. For vacuum inputs  we obtain
\begin{align}
\frac{d}{dt}\rho_{\rm sys}&=-\frac{i}{\hbar}[H_{\rm sys},\rho_{\rm sys}]+\gamma(\mathcal{D}[c_1]\rho_{\rm sys}+\mathcal{D}[c_2]\rho_{\rm sys})\no\\
&-(\gamma_Le^{i\phi}[c_1,\rho_{\rm sys} c_2^{\dag}]+\!\gamma_Re^{i\phi}[c_2,\rho_{\rm sys} c_1^{\dag}]\!-\!\textrm{h.c.}),\label{MasterEquation}
\end{align}
with $\mathcal{D}[c]\rho\equiv c\rho c^\dag-\frac{1}{2}\{c^\dag c,\rho\}$ \cite{Carmichael:1993el,Gardiner:1993cy}. We note that Eq.~\eqref{MasterEquation} contains an {\em instantaneous} dipole-dipole interactions, and {\em collective} atomic decay terms related to the 1D character of the reservoir. For the case of symmetric decay, $\gamma_L=\gamma_R$, \eqref{MasterEquation} describes  super- and subradiant decay processes \cite{Chang:2012wb}. In the limit of {\em purely unidirectional couplings}, $\gamma_L=0$, where node $1$ drives node $2$, but there is no back scattering from $2$ to $1$, the above equation reduces to the master equation for a {\em cascaded quantum system}, as first derived by Carmichael and Gardiner \cite{Gardiner:1993cy,Carmichael:1993el}. We note that for the cascaded case a finite time delay $\tau>0$ can always be absorbed in a retarded time for node 2, i.e.~the system dynamics can be described by a Markovian master equation \cite{QWI,QWII}. This is not the case, however, when we allow for back scattering or two-way communication. Below we address this problem by solving the QSSE for $\tau>0$, where a Markovian master equation of the type \eqref{MasterEquation} does not exist.  

To give a meaning to a QSSE with time delays, and to prepare our MPS
formulation to its solution we find it convenient to discretize time
in small steps $\Delta t$, that is $t_{k}=k\Delta t$ with $k\in\mathbb{Z}$.
We represent the time evolution as a dynamical map $\ket{\Psi(t_{k+1})}=U_{k}\ket{\Psi(t_{k})}$.
We choose a time step, which is small compared the timescale of the
system evolution (including $\gamma_{L,R}\Delta t\ll 1$), but large compared
to the inverse bandwidth $\cal B$ of the waveguide.  Moreover, we conveniently
choose $\Delta t$ a unit fraction of the delay time, $\tau=\ell\Delta t$
(with $\ell$ integer). Thus we have 
\begin{align}
\ket{\Psi(t_{k+1})} & =U_{k}\ket{\Psi(t_{k})}\label{eq:QSSEmap}\\
& \equiv\exp\lr{-\frac{i}{\hbar}{H}_{{\rm sys}}(t_{k})\Delta t+O_{k,1}+O_{k,2}}\ket{\Psi(t_{k})}\nonumber
\end{align}
with 
\begin{align}
\begin{split}O_{k,1}\! & =\!\big(\sqrt{\gamma_{L}}\Delta B_{L}^{\dag}\!\lr{t_{k}}+\!\sqrt{\gamma_{R}}\Delta B_{R}^{\dag}\!\lr{t_{k-\ell}}\!e^{i\phi}\big)c_{1}\!-\!\textrm{h.c.}\label{eq:QSSEOp1}\\
O_{k,2}\! & =\!\big(\sqrt{\gamma_{L}}\Delta B_{L}^{\dag}\!\lr{t_{k-\ell}}\!e^{i\phi}+\!\sqrt{\gamma_{R}}\Delta B_{R}^{\dag}\!\lr{t_{k}}\!\big)c_{2}\!-\!\textrm{h.c.}
\end{split}
\end{align}
Here we have defined quantum noise increments $\Delta B_{i}(t_{k})=\int_{t_{k}}^{t_{k+1}}dt\,b_{i}(t)$.
They obey (up to a normalization factor) bosonic commutation relations,
$[\Delta B_{i}(t_{k}),\Delta B_{i'}^{\dag}(t_{k'})]=\Delta t\delta_{i,i'}\delta_{k,k'}$,
and can thus be interpreted as annihilation (or creation) operators
for photons in the time bin $k$. Since we have two channels ($L,R$)
each time bin contains two such modes.  The above equation states that
in time step $t_{k}\rightarrow t_{k+1}$ the first node can emit a
photon into the two modes, the $L$-mode of bin $k$ and the $R$-mode
of bin $k-\ell$, and vice versa for the second node. Thus we can visualize the time evolution as a {\em conveyor belt} of time bins representing the modes (c.f.~Fig 1(c)): each time step shifts this conveyor belt by one unit, and after $\ell$ such steps the first (second) system interacts with the photons emitted by the second (first) one. 

\textit{Matrix Product State description.} In the following we employ a MPS representation of $\ket{\Psi(t)}$.  In a condensed matter context time-dependent Density Matrix Renormalization Group (tDMRG) techniques have been developed to integrate the many-particle Schr\"odinger equation in 1D systems, and ladder geometries, and 
a close relationship between MPS and the output fields from photonic systems has been established \cite{Schon:2005fk,Schon:2007bw,Osborne:2010hr,Verstraete:2010bf,Haegeman:2010bc}. Here we build on these developments to integrate \eqref{eq:QSSEOp1} efficiently and for long times, approaching the steady state.

We assume that the full state is initially ($t=0$) completely
uncorrelated, that is $\ket{\Psi(t=0)}=\ket{\psi_{S}}\bigotimes_{p}\ket{\phi_{p}}$,
where $\ket{\psi_{S}}$ denotes the initial state of the nodes (emitters) 
and $\ket{\phi_{p}}$ the state of the photons in time bin $p$. In
particular this includes a waveguide initially in the vacuum state
$\ket{\phi_{p}}=\ket{\mbox{\rm vac}_p}$ with $\Delta B_{L/R}(t_p)\ket{\rm vac_p}=0$. The stroboscopic evolution
until a time $t_{k}$ gives a state $\ket{\Psi(t_{k})}=\ket{\phi_{in}(t_{k})}\otimes\ket{\psi(t_{k})}$,
where $\ket{\phi_{in}(t_{k})}=\bigotimes_{p\geq k}\ket{\phi_{p}}$
is the remaining (unused) input state and 
\begin{align}\label{eq:state}
\ket{\psi(t_{k})} & =\!\!\!\!\!\sum_{i_{S},\{{i_{p}}\}}\!\!\!\!\psi_{i_{S},i_{k-1},i_{k-2},\dots}\ket{i_{S},i_{k-1},i_{k-2}\dots}
\end{align}
is the entangled state of nodes and radiation field. Here $i_{S}$
and $i_{p}$ label the basis states in the Hilbert space of the
nodes and the time bin $p$, respectively:
\begin{align}
\ket{i_p}\equiv \ket{i_p^L,i_p^{R}}=\frac{(\Delta B_L^{\dag})^{i_p^L}}{\sqrt{\Delta t ^{i_p^L} i_q^L!}}\frac{(\Delta B_R^{\dag})^{i_p^R}}{\sqrt{\Delta t ^{i_p^R} i_q^R!}}\ket{{\rm vac}_p}
\end{align}
where $i_p^{L/R}=0,1,2\dots$ denote the number of photons in the $L/R$ mode in time bin $p$.
The modes
$p\in(k-1,\dots,k-\ell)$ represent the photonic state in the quantum
circuit, while $p<k-\ell$ labels the modes of the output field {[}cf.~Fig.~\ref{Fig1}(b){]}.

In a MPS form \cite{White:1992zz,Vidal:2004jc,Daley:2004hk,Schollwock:2005wz,Schollwock:2011gl}
we can write these amplitudes as
\begin{align}
\psi_{i_{S},{i}_{k-1},\dots\!}=\tr{A[S]^{i_{S}}A[k-\!1]^{i_{k-1}}A[k-\!2]^{i_{k-2}}\!\!\dots},\label{eq:MPS}
\end{align}
where $A[p]^{i_{p}}$ are $D_{p}\times D_{p-1}$ matrices and
$D_{p}$ is the bond dimension for a bipartite cut between time bins
$p$ and $p+1$. We emphasize that the quantum state \eqref{eq:state}, and its MPS decomposition \eqref{eq:MPS}, refer to an entangled state of nodes and photons {\em in time bins}. This is in contrast to condensed matter systems where the many-body wavefunction refers to {\em spatial}  correlations at a given time. 
The propagation of the full state from $t_{k}$ to $t_{k+1}$ via
Eq.~\eqref{eq:QSSEOp1} involves an interaction of each node with
two time-bin modes of the radiation fields, with time delays appearing as ``long range interactions''. While instantaneous (short range) interactions are standard to implement in the MPS formalism \cite{Vidal:2004jc,Daley:2004hk},
time-delayed (long range) interactions can be handled by methods introduced in \cite{Schachenmayer:2010ia}.
In each time step the MPS \eqref{eq:MPS} grows by one
site ($A[k]$). For the technical details for updating the MPS state in each time step we refer to the Supplemental Material \cite{SUPP}. Note however that only the state of the nodes and the time bins in $[t_k-\tau,t_k]$ are
updated in the $k$th time step.

\begin{figure}[h]
\centering \includegraphics[width=0.5\textwidth]{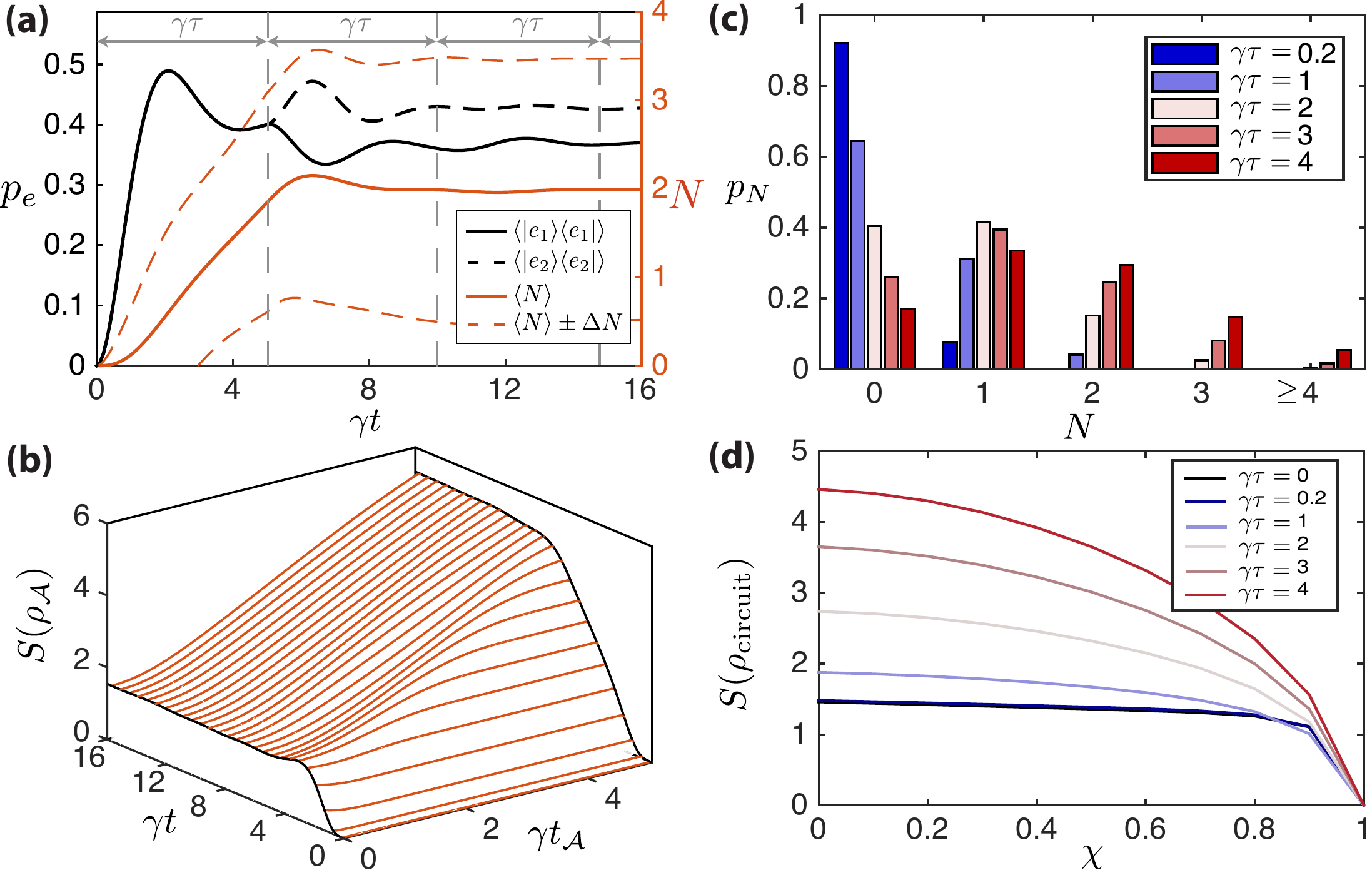} \caption{Two driven two-level atoms coupled to a waveguide [c.f.~Fig.~\ref{Fig1}(c)]. (a) Atomic
excitation probabilities of atoms 1 and 2 (solid/dashed black lines), and photon number in the waveguide between the atoms (red) as function of time for $\gamma\tau=5$ and $|\Omega|=1.5\gamma$.  Vertical dashed lines indicate multiples of the delay time.
(b) Entanglement entropy $S(\rho_{\mathcal{A}})$ of the atoms and radiation field in the interval $[t-t_\mathcal{A},t]$ as a function of time. The line at $t_{\mathcal{A}}=0$ gives the entanglement
of the atoms with the entire radiation field, while the line at $t_{\mathcal{A}}=\tau=5/\gamma$
corresponds to the entanglement of the entire circuit with the output field. (c) Probabilities $p_{N}$ for
having $N$ photons in the waveguide between the atoms for different delay times
$\tau$ in the steady state (calculated by time evolution to $t_{\rm max}=200/\gamma$). 
(d) Entanglement entropy of the entire circuit with the output field in the steady
state for asymmetric coupling $\gamma_{L/R}=\gamma(1\pm \chi)/2$. Unless otherwise stated parameters in (a-d) are:  $\gamma_L=\gamma_R\equiv\gamma/2$, $\gamma\tau=5$, $\phi=\pi/2$, $\Omega_1=\Omega_2 e^{i\phi}=1.5\gamma$, $\Delta=0$; $\gamma\Delta t=0.1$, $D_{{\rm max}}=256$.}
\label{Fig2} 
\end{figure}

We now illustrate this generic method with two examples: (i) two coherently driven, distant atoms interacting via the waveguide as discussed above, and (ii) a single driven atom coupled to a waveguide terminated by a distant mirror as illustration of a quantum feedback problem. In contrast to previous work \cite{LeKien:2005bo,Bushev:2006ip,GonzalezTudela:2011fs,Chang:2012wb,Milonni:1974bo,Dorner:2002dv,Buzek:1999hk,GieSen:1996ct,Feng:1990gw,Cook:1987fr,Rist:2008cm,Glaetzle:2010dt,Zheng:2013fy,Fang:2015bc,Zeeb:2015kz}, which includes transfer matrix and Wigner-Weisskopf type approaches applicable to a single, or a few excitations propagating through the system, we are interested here in strongly driven systems with multiple photon exchange resulting in significant entanglement. The problem of delayed quantum feedback was addressed recently by Grimsmo \cite{Grimsmo:2015gf} in the transient regime of a few delay cycles. In contrast we will be able to follow the evolution of the circuit for long times, reaching the steady state. 

\textit{Two driven distant atoms.} Figs.~\ref{Fig2}(a,b) show results for the time evolution of two atoms driven through the waveguide from the left input port, which are separated by a distance corresponding to a delay $\gamma\tau=5$, and a propagation phase $\phi=\pi/2$ [see Fig.~\ref{Fig1}(c)]. Fig.~\ref{Fig2}(a) plots the atomic excitation probabilities and the mean photon number in the waveguide between the two atoms (delay line). For a time evolution starting at $t=0$, the atoms will ``not see each other'' for times $0\le t < \tau$, and thus obey Rabi dynamics described by single atom Bloch equations. For $t>\tau$ the atom interacts both with the coherent drive and also with the time-delayed non-classical stream of photons emitted by the other atom. We assume a driving laser field from the left and thus the output field of atom $1$ acquires the same phase as the laser, when traveling to atom $2$. On the other hand the output field of the atom $2$ is out of phase with respect to the coherent drive by $2\phi$ when it reaches atom $1$. This causes the destructive and constructive interference in the atomic populations for $t>\tau$ in Fig.~\ref{Fig2}(a).

\begin{figure}[t!]
\centering \includegraphics[width=0.5\textwidth]{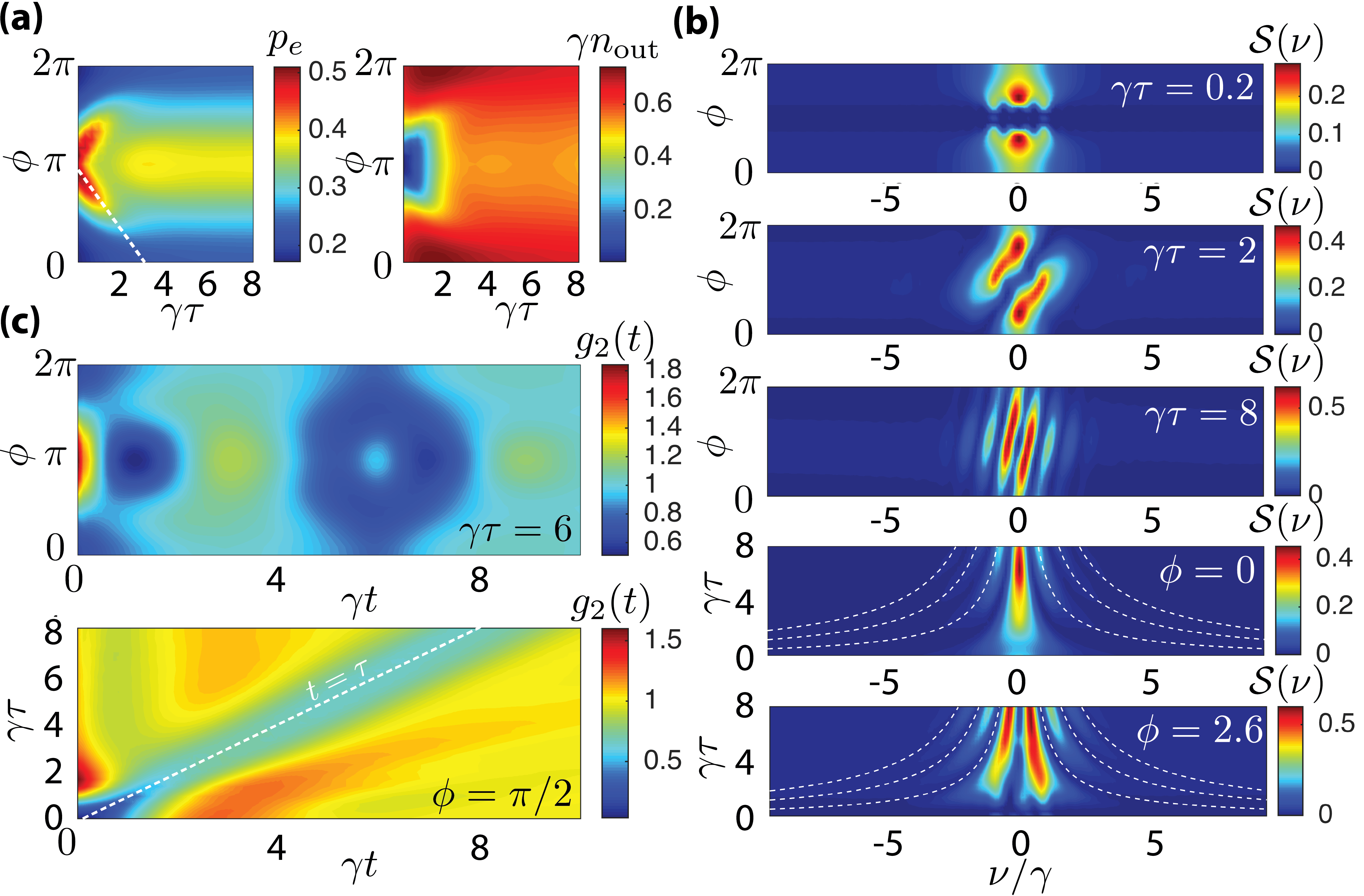} \caption{Steady state properties of the circuit and output field for different delays $\gamma\tau$ and feedback
phases $\phi$ (with $\gamma_L=\gamma_R=\gamma/2$). (a) Excited state probabilities
of the atom (left) and intensity of the output photo-current (right). The
white line corresponds to $\phi=\pi-\sqrt{\Omega^{2}+\Delta^{2}}\tau$,
where the feedback is out of phase with the drive. (b) Incoherent
part of the spectrum of the output field as a function of the feedback
phase for $\gamma\tau=0.2,\,2,\,4$ and as function of the delay time
for $\phi=0,\,2.6$ (from top to bottom). (c) Autocorrelation function
of the output field for $\gamma\tau=6$ (top) and for different delay
times at $\phi=\pi/2$ (bottom). Parameters: $\gamma\Delta t=0.1$, $D_{{\rm {max}}}=50$). The steady state is obtained by a evolution up to $t_{{\rm end}}=200/\gamma$.}
\label{Fig4} 
\end{figure}

In Fig.~\ref{Fig2}(b) we plot the time evolution of the entanglement of time bins in terms of the entropy $S(\rho_{\mathcal{A}})\equiv-\textrm{Tr}\{\rho_{\mathcal{A}}\log_{2}\rho_{\mathcal{A}}\}$ of the
reduced state $\rho_{\mathcal{A}}(t)=\textrm{Tr}_{\neg\mathcal{A}}\{\ket{\Psi(t)}\bra{\Psi(t)}\}$. Here $\mathcal{A}$ refers to the radiation field in time bins $[t-t_\mathcal{A},t]$ [Fig.~\ref{Fig1}(d)]. Note that for $t_\mathcal{A}=0$ the state $\rho_\mathcal{A}\equiv\rho_{\rm sys}$ corresponds to the reduced density operator of the atoms, and $S(\rho_{\rm sys})$ quantifies the atom-photon entanglement, as the total state $\ket{\Psi(t)}$ is pure. On the other hand, for $t_\mathcal{A}=\tau$, the state $\rho_\mathcal{A}\equiv\rho_{\rm circuit}$ includes also the radiation field in the waveguide between the atoms, and thus $S(\rho_{\rm circuit})$ quantifies the entanglement of the circuit with the output field \footnote{For $t_\mathcal{A}>\tau$, the entanglement is per construction $S(\rho_\mathcal{A}(t))=S(\rho_{\rm circuit}(t-t_\mathcal{A}+\tau))$. Thus $S(\rho_{\rm circuit})$ sets the maximum entanglement generated in the MPS.}.  $S(\rho_{\rm circuit})$ increases approximately linearly during the first roundtrip time, but does not increase afterwards. The necessary bond dimension to represent the state, $D_{\rm max}$ scales thus exponentially with $\gamma\tau$, which limits the achievable time delays. However for a fixed $\tau$, the bond dimension does not increase with the total integration time, allowing us to reach the steady state. This can be understood by noting that each photon is emitted as a superposition state into the left and right moving channel, and contributes an entropy $S_1=-(\gamma_L/\gamma)\log_2(\gamma_L/\gamma)-(\gamma_R/\gamma)\log_2(\gamma_R/\gamma)$ to the total entanglement of the circuit for a time $\tau$ after its emission, i.e.~before the photon leaves the circuit \cite{SUPP}. $S(\rho_{\rm circuit})$ thus scales linearly with the number of photons present in the waveguide between the atoms \footnote{We note that the entanglement does not depend on the choice of the time step $\Delta t$.}.   In Fig.~\ref{Fig2}(c) we plot this photon number distribution in steady state for increasing $\tau$. Fig.~\ref{Fig2}(d) shows the corresponding increase of the entanglement $S(\rho_{\rm circuit})$, and the role of chirality $\gamma_L\neq\gamma_R$.
The entropy per photon $S_1$ is maximal for a bidirectional system $(\gamma_L=\gamma_R$), and vanishes in the unidirectional case ($\gamma_R=0$), such that in the cascaded limit $S(\rho_{\rm circuit})$ becomes independent of $\tau$. This connects our approach to the well known fact that time delays can be trivially eliminated in the purely cascaded limit (see above and \cite{Carmichael:1993el,Gardiner:1993cy}).  The vanishing of the entropy $S(\rho_{\rm circuit})$ in Fig.~\ref{Fig2}(d) for the purely unidirectional coupling $\gamma_R=0$ indicates the existence of a dark (pure) quantum state as steady state of the quantum circuit. The formation as quantum dimers of two-level atoms has been discussed in the Markovian case \cite{Stannigel:2012jk}, and these persist as dimer correlations shifted by the time delay between the two atoms even for $\tau>0$ \cite{RamosVermersch}.

\textit{Quantum feedback -- atom in front of a mirror:} Finally we illustrate our approach for the example of a driven atom in front of a mirror at distance $d$ (see also \cite{Grimsmo:2015gf}), and  calculate properties of the atomic steady state and the corresponding output field (Fig.~\ref{Fig4}). The quantum stochastic Hamiltonian is given by \cite{SUPP}
 \begin{align}
{H}_{{\rm int}}(t)\!=&i\hbar(\sqrt{\gamma_{R}}b^{\dag}\!\lr{t}\!+\!\sqrt{\gamma_{L}}b^{\dag}\!\lr{t\!-\!\tau}e^{i\phi})\ket{g}\!\bra{e}\!-\!\textrm{h.c.})\label{Eq2-2}
\end{align} 
The parameters characterizing this setup are the delaytime $\tau=2d/v$ and the roundtrip propagation phase $\phi=\pi-\bar\omega \tau$. In the Markovian limit, $\tau\rightarrow 0^+$, the system is described by the well known optical Bloch equation, with effective decay rate $\gamma_{\rm eff}=2\gamma\cos^ 2(\phi/2)$ and effective detuning $\Delta_{\rm eff}=\Delta-(\gamma/2)\sin(\phi)$ (for $\gamma_L=\gamma_R$) \cite{Horak:2010js,Eschner:2001ib,Wilson:2003ef,Beige:2002gf}. In this limit the output power spectrum in the steady state, $\mathcal{S}(\nu)$, shows a Mollow triplet \cite{Mollow:1969wc} and the autocorrelaltion function $g_2(t)$ exhibits photon antibunching. With our methods we can go systematically beyond this limit and calculate these steady state quantities for long delays $\tau\gg\gamma^{-1},\Omega^{-1}$ (Fig.~\ref{Fig4}). As depicted in Fig.~\ref{Fig4}(b) with increasing $\tau$ the incoherent part of the spectrum develops a series of peaks at $\nu=(\phi+2\pi\mathbb{Z})/\tau$. This reflects the coherence of photons that are emitted in a superposition of states corresponding to propagation towards and away from the mirror, resulting in correlations of time bins separated by $\tau$. As shown in Fig.~\ref{Fig4}(c) $g_2(t)$ also reveals long time correlations $g_2(\tau)<1$. This reduced probability of detecting two photons delayed by $\tau$ can be traced back to the anti-bunching of photons emitted towards and away from the mirror. Moreover, depending on the phase of the feedback $\phi$, the equal time autocorrelation function $g_2(t=0)$ can change from the well known antibunching dip $(g_2(0)<1)$ to a bunching peak $(g_2(0)>1)$, where photons in the feedback line interfere with the emission of  photons directly into the output port. [Fig.~\ref{Fig4}(c)]. 

In summary, we have developed a MPS approach to describe the dynamics of photonic quantum networks with time delays, and the quantum feedback problem. This provides a systematic framework to simulate complex, composite nonlinear photonic quantum circuits and quantum optical devices with several input and output channels \cite{SUPP}.

\textit{Acknowledgment:} We thank H. Carmichael, C. W. Gardiner,  A. M. L\"auchli, T. Ramos and B. Vermersch for discussions. Work at Innsbruck is supported by the ERC Synergy Grant UQUAM, the Austrian Science Fund through SFB FOQUS, and EU FET Proactive Initiative SIQS. The authors thank the Solvay Institute Brussels for hospitality.

 \bibliographystyle{apsrev4-1.bst}
\bibliography{HannesBibTex,SUPP_bibtex,Additional_refs}

\newpage
\newpage 

\onecolumngrid
\newpage
{
\center \bf \large 
Supplemental Material for: \\
Photonic Quantum Circuits with Time Delays: A Matrix Product State Approach\vspace*{0.1cm}\\ 
\vspace*{0.0cm}
}
\begin{center}
Hannes Pichler$^{1,2}$ and Peter Zoller$^{1,2}$\\
\vspace*{0.15cm}
\small{\textit{$^1$Institute for Quantum Optics and Quantum Information of the Austrian
Academy of Sciences, 6020 Innsbruck, Austria\\
$^2$Institute for Theoretical Physics, University of Innsbruck, 6020 Innsbruck, Austria}}\\
\vspace*{0.25cm}
\end{center}

\twocolumngrid

\section{Derivation of the Quantum Stochastic Schr\"odinger Equation}

Here we summarize briefly the derivation of the QSSE (see Ch.~9 of \cite{QWII}) for the two examples considered in the main text, namely two atoms interacting via a 1D waveguide with Hamiltonian Eq.~(\ref{Eq2-1}), and a single atom coupled to a waveguide terminated by a mirror according to Eq.~(\ref{Eq2-2}).

\subsection{Setup 1: two atoms coupled to a waveguide}\label{SUPP_Two_atoms}

We consider two atoms ($n=1,2$) representing two nodes of the quantum network, which we model as two-level atoms with ground and excited states $\ket{g_n}$ and $\ket{e_n}$, respectively, coupled to a 1D waveguide with left and right propagating modes (see Fig.~\ref{FigSUPP1}(a)).  The Schr\"odinger equation for the total state of atoms and waveguide $\ket{\Psi(t)}$ is given by 
\[
i\hbar\frac{d}{dt}\ket{\Psi(t)}=H_{\rm tot}\ket{\Psi(t)}
\]
with total Hamiltonian
\[
H_{{\rm tot}}=H_{{\rm sys}}+H_{B}+H_{{\rm int}}.
\]
Here, $H_{{\rm sys}}$ refers to the bare dynamics of the
two atoms
\begin{align}\label{SUPP1}
H_{{\rm sys}}=\sum_{n=1,2}\left[ \hbar\omega_{n}\ket{e_n}\bra{e_n}-\frac{\hbar}{2}\lr{\Omega_n\ket{g_n}\bra{e_n}e^{i\bar \omega t}+\rm h.c.}\right].
\end{align}
Here $\omega_{n}$ denotes the transition frequency of the atom $n$. The atoms are driven by a coherent driving field with Rabi frequency $\Omega_n$ at a frequency $\bar \omega$, detuned by $\Delta _n=\bar \omega-\omega_n$ from the atomic transition frequency. Note that in writing the atomic Hamiltonian in the form \eqref{SUPP1} we made a rotating wave approximation valid for $|\Delta_n|,|\Omega_n|\ll\bar \omega$.
The bare Hamiltonian of the waveguide is given by 
\begin{align}\label{SUPP2}
H_{B}=\sum_{i=L,R}\int_{\mathcal{B}}d\omega\hbar\omega b_{i}^{\dag}(\omega)b_{i}(\omega),
\end{align}
where the bosonic operators $b_i(\omega)$ ($b_i^\dag(\omega)$) destroy (create) a photon of frequency $\omega$ propagating to the left ($i=L$) or to the right ($i=R$). They obey the usual bosonic commutation relations $[b_{i}(\omega),b_{i'}^{\dag}(\omega')]=\delta_{i,i'}\delta(\omega-\omega')$. 

The interaction between each atom and light reads in the RWA
\begin{align}\label{SUPP3}
H_{\rm int}=i\hbar\sum_{i,n}\int_{\mathcal{B}} d\omega\, \kappa_i(\omega) [b_{i}^\dag(\omega)c_{n} e^{-i\omega x_n/v_i}-\rm{h.c.}],
\end{align}
where $x_n$ denotes the position of the atom $n$ along the waveguide, and $v_i$ the photon group velocity. We conveniently  choose $x_2>x_1$ and $v_R=-v_L\equiv v>0$.  For consistency with the rotating wave approximations we include only photons in the relevant bandwidth $\mathcal{B}$ around $\bar\omega$ in \eqref{SUPP2} and \eqref{SUPP3}. The transition operators associated with the atomic decay are denoted by $c_{n}=\ket{g_n}\bra{e_n}$, and the coupling matrix elements for creating a left or right ($i=L,R$) moving photon of frequency $\omega$ are $\kappa_i(\omega)$. The position of the atom $x_n$ enters here as a phase factor $e^{-i\omega x_n/v_i}$. We note that due to the presence of more than one atom this phase factor can not simply be gauged away. Following the standard quantum optical treatment \cite{QWI,QWII} we assume the modulus of the coupling elements to be constant over the relevant bandwith $\mathcal{B}$, and approximate $\kappa_i(\omega)\rightarrow \sqrt{{\gamma_i}/{2\pi}}$. The parameters  $\gamma_L$ and $\gamma_R$ correspond to the single atom decay rates into left and right moving photons, respectively. For notational simplicity we assumed them to be the same for both atoms.

\subsubsection{Quantum Stochastic Schrödinger equation}

To derive a QSSE we go into an interaction picture with respect to the bath Hamiltonian $H_B$, and into a rotating frame with the frequency of the driving laser. The corresponding unitary transformation is 
\begin{align}\label{SUPP_UI}
U_I(t)=\exp\lr{-i H_Bt/\hbar-i\sum_n\bar\omega\ket{e_n}\bra{e_n} t}.
\end{align}
In this interaction picture the Schr\"odinger equation for the state $\ket{\Psi_I(t)}=U_I^{\dag}(t)\ket{\Psi(t)}$ reads 
\begin{align}
i\hbar\frac{d}{dt}\ket{\Psi_I(t)}=(H_{{\rm sys}, I}+H_{{\rm int},I}(t))\ket{\Psi_I(t)},
\end{align}
where the Hamiltonian in the interaction picture consists again of a system part 
\begin{align}\label{SUPP_HsysI}
H_{{\rm sys}, I}=\sum_n \lr{-\hbar\Delta_{n}\ket{e_n}\bra{e_n}-\frac{\hbar}{2}\lr{\Omega_n\ket{g_n}\bra{e_n}+\rm h.c.}},
\end{align}
and an interaction part 
\begin{align}
{H}_{{\rm int}, I}(t) & \!=\!i\hbar\!\sum_{i,n}\left(\sqrt{\gamma_{i}}b_{i}^{\dag}\!\lr{t-x_n/v_i}e^{-i\bar\omega x_n/v_i}c_{n}\!-\!\textrm{h.c.}\!\right).
\end{align}
Here we have defined quantum noise operators \begin{align}\label{SUPP_noise}
b_i(t)=\frac{1}{\sqrt{2\pi}}\int_\mathcal{B} d\omega b_i(\omega)e^{-i(\omega-\bar \omega )t},
\end{align}
obeying the white noise commutation relations $[b_i(t),b_i^\dag (t')]=\delta_{i,j}\delta(t-t')$. In going to the interaction picture we thus essentially changed from a {\em frequency} representation of the radiation field to a {\em time} representation.
To obtain a complete formal equivalence with the form of the QSSE \eqref{Eq2-1} given in the main text we perform a variable shift and redefine the phase of this noise operators by 
\begin{align}
b_L(t)&\rightarrow b_L(t+x_1/v_L)e^{-i\bar\omega x_1/v_L},\\
b_R(t)&\rightarrow b_R(t+x_2/v_R)e^{-i\bar\omega x_2/v_R}.
\end{align}
With this and the identification $\tau\equiv (x_2-x_1)/v$ as well as $\phi=-\bar \omega \tau$, we recover the QSSE with the Hamiltonian \eqref{Eq2-1} of the main text:
\begin{align}
{H}_{{\rm int},I}(t) & =i\hbar\left(\big(\sqrt{\gamma_{L}}b_{L}^{\dag}\lr{t}+\sqrt{\gamma_{R}}b_{R}^{\dag}\lr{t-\tau}e^{i\phi}\big)c_{1}-\textrm{h.c.}\right)\no\\
& +i\hbar\left(\big(\sqrt{\gamma_{L}}b_{L}^{\dag}\lr{t-\tau}e^{i\phi}+\sqrt{\gamma_{R}}b_{R}^{\dag}\lr{t}\big)c_{2}-\textrm{h.c.}\right).\no
\end{align}
In the main text we drop the subscript $I$.

\subsubsection{Discretization into time bins}
As discussed in the main text we interpret this QSSE in terms of a stroboscopic map with discrete time steps $\Delta t$. In this time bin representation we discretize time in steps, that is $t_0=0$ and $t_{k+1}=t_k+\Delta t$ with $k\in \mathbb{Z}$. The time step $\Delta t$ has to be chosen much smaller that the system timescales associated with the coherent drive $|\Omega_n|$, its detuning $|\Delta_n|$, and the decay rates into the waveguide $\gamma_{L,R}$. 
\begin{align}
\Delta t\ll \gamma_{L,R}^{-1},|\Omega_{1,2}|^{-1},|\Delta_{1,2}|^{-1}
\end{align}
We note that $\Delta t$ should be larger that the timescale associated with the bandwidth $\mathcal{B}$ of the photon modes that we kept in the RWA. In practice this is not a limitation and we formally let $\Delta t\rightarrow 0^{+}$. Thus, for a finite delay $\tau$ we can always choose a $\Delta t$ such that it is an integer fraction of $\tau$.

For each of these time bins we define quantum noise increments $$\Delta B_{i}(t_{k})=\int_{t_{k}}^{t_{k+1}}dt\,b_{i}(t).$$
that obey (up to a normalization factor) bosonic commutation relations,
$$[\Delta B_{i}(t_{k}),\Delta B_{i'}^{\dag}(t_{k'})]=\Delta t\delta_{i,i'}\delta_{k,k'}.$$
and can thus be interpreted as annihilation (or creation) operators
for photons in the time bin $k$. 

\begin{figure}
\centering \includegraphics[width=\linewidth]{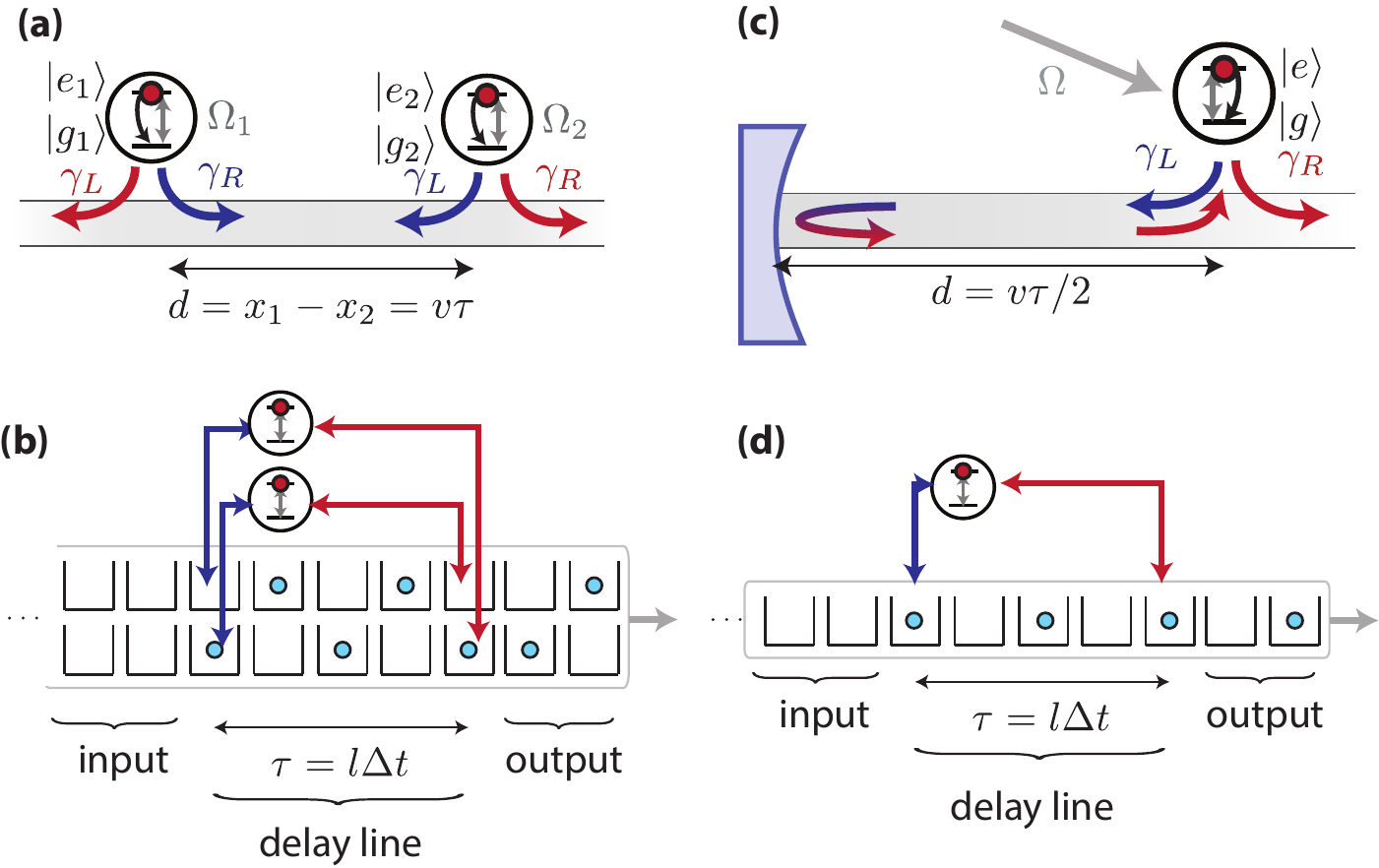} \caption{
(a,b) Setup 1 considered in the main text: (a) two atoms  at a distance $d$ interacting with a waveguide. (b) Time bin representation of the QSSE as a stroboscopic map for the evolution of the atoms and the radiation fields in a single time step. (c,d) Setup 2 considered in the main text (quantum feedback): (c) atom in front of a distant mirror. (d) Time bin representation of the QSSE, and interpretation of the stroboscopic map describing the evolution of the atoms and the radiation fields for a single time step.}
\label{FigSUPP1} 
\end{figure}

With this we can represent the time evolution according to the QSSE to lowest order in $\Delta t$ as the dynamical map stated in the main text:
\begin{align}
\ket{\Psi_I(t_{k+1})} & =U_{k}\ket{\Psi_I(t_{k})}\label{SUPP_QSSEmap}\\
& \equiv\exp\lr{-\frac{i}{\hbar}{H}_{{\rm sys}}(t_{k})\Delta t+O_{k,1}+O_{k,2}}\ket{\Psi_I(t_{k})}\nonumber
\end{align}
with 
\begin{align}
\begin{split}O_{k,1}\! & =\!\big(\sqrt{\gamma_{L}}\Delta B_{L}^{\dag}\!\lr{t_{k}}+\!\sqrt{\gamma_{R}}\Delta B_{R}^{\dag}\!\lr{t_{k-\ell}}\!e^{i\phi}\big)c_{1}\!-\!\textrm{h.c.}\label{SUPP_QSSEOp1}\\
O_{k,2}\! & =\!\big(\sqrt{\gamma_{L}}\Delta B_{L}^{\dag}\!\lr{t_{k-\ell}}\!e^{i\phi}+\!\sqrt{\gamma_{R}}\Delta B_{R}^{\dag}\!\lr{t_{k}}\!\big)c_{2}\!-\!\textrm{h.c.}
\end{split}
\end{align}
Fig.~\ref{FigSUPP1}(b) is a  graphical representation of the {\em time bin representation}, illustrating the propagation for a single time step.

\subsection{Setup 2: single atom in front of a mirror}\label{SUPP_feedback}

The second example studied in the main text considers the problem of a single atom coupled to a waveguide terminated at distance $d$ by a mirror, i.e.~a half cavity (see Fig.~\ref{FigSUPP_Time}(c))). Similar to the previous example the Hamiltonian of the combined system is of the form $H_{{\rm tot}}=H_{{\rm sys}}+H_{B}+H_{{\rm int}}$. The system Hamiltonian is analogous to $\eqref{SUPP1}$ for a single atom 
\begin{align}\label{SUPPmirror1}
H_{{\rm sys}}=\hbar\omega\ket{e}\bra{e}-\frac{\hbar}{2}\lr{\Omega\ket{g}\bra{e}e^{i\bar \omega t}+\rm h.c.}.
\end{align}
In contrast to the $i=L,R$ modes of the previous example, we have here only one mode, so that the Hamiltonian for the radiation field reads
\begin{align}\label{SUPPmirror2}
H_{B}=\int_{\mathcal{B}}d\omega\hbar\omega b^{\dag}(\omega)b(\omega),
\end{align}
where the bosonic operators $b(\omega)$ ($b^\dag(\omega)$) again destroy (create) a photon of frequency $\omega$. The corresponding mode functions are superpositions of plane waves with positive and negative wavevector satisfying the boundary condition of zero electric field at the position of the mirror, chosen at $x=0$.
The atom-photon interaction of the atom placed at a distance $d$ from the mirror is then given by
\begin{align}\label{SUPPmirror3}
H_{\rm int}\!=i\hbar\!\int_{\mathcal{B}} \!\!d\omega\,  [b^\dag(\omega)c (\kappa_R(\omega)e^{-i\omega d/v}\!-\!\kappa_L(\omega)e^{i\omega d/v})\!-\!\rm{h.c.}],\no
\end{align}
with $c=\ket{g}\bra{e}$.
\subsubsection{QSSE for an atom in front of a mirror}
Similar to \eqref{SUPP_UI} we go to an interaction picture
\begin{align}
i\hbar\frac{d}{dt}\ket{\Psi_I(t)}=(H_{{\rm sys}, I}+H_{{\rm int},I}(t))\ket{\Psi_I(t)},
\end{align}
where $H_{{\rm sys}, I}$ is (compare Eq.~\eqref{SUPP_HsysI})
and
\begin{align}
{H}_{{\rm int}, I}(t)&=i\hbar\sqrt{\gamma_{R}}b^{\dag}\lr{t-d/v}e^{-i\bar\omega d/v}c\no\\
&-i\hbar\sqrt{\gamma_{L}}b^{\dag}\lr{t+d/v}e^{+i\bar\omega d/v}c+\textrm{h.c.}
\end{align}
with quantum noise operators defined as in \eqref{SUPP_noise}. By identifying the roundtrip time $\tau=2d/v$ and the phase $\phi=\pi-\bar\omega\tau$, we can redefine 
\begin{align}
b(t)\rightarrow b(t+\tau/2)e^{i\phi/2}
\end{align}
and write the interaction Hamiltonian in the form used in the main text
\begin{align}
{H}_{{\rm int}, I}(t)=&i\hbar(\sqrt{\gamma_{R}}b^{\dag}\lr{t}+\sqrt{\gamma_{L}}b^{\dag}\lr{t-\tau}e^{i\phi})c-\textrm{h.c.}).\no
\end{align}
In the main text we drop the subscript $I$.

\subsubsection{Discretized version}

By discretizing time in small steps $\Delta t$ we obtain the stroboscopic map (illustrated in Fig.~\ref{FigSUPP_Time}(d))) 
\begin{align}
\ket{\Psi_I(t_{k+1})} & =U_{k}\ket{\Psi_I(t_{k})}\label{SUPP_Uk}\\
& \equiv\exp\lr{-\frac{i}{\hbar}{H}_{{\rm sys},I}(t_{k})\Delta t+O_{k}}\ket{\Psi_I(t_{k})}\no
\end{align}
with
\begin{align}
O_{k} =\big(\sqrt{\gamma_{R}}\Delta B^{\dag}\!\lr{t_{k}}+\sqrt{\gamma_{L}}\Delta B^{\dag}\!\lr{t_{k-\ell}}\!e^{i\phi}\big)c\!-\!\textrm{h.c.},
\end{align}
that propagates the state in an analogous way as described in the main text for the situation of two systems coupled to the same waveguide in \eqref{eq:QSSEOp1}. Here the $\Delta B(t_k)=\int_{t_k}^{t_{k+1}}dt b(t)$ are bosonic operators ($[\Delta B(t_k),\Delta B^\dag(t_{k'})]=\Delta t \delta_{k,k'}$) that annihilate a photon in the time bin $k$. 

For a discussion of the corresponding dynamics we refer to Fig.~\ref{FigSUPP_Time}.

\section{State evolution and Matrix Product state update}

Here we elaborate on the propagation of the state and the corresponding update MPS
for a single time step $t_{k}\rightarrow t_{k+1}$ for a stroboscopic evolution with $U_k$ of the form \eqref{SUPP_QSSEmap} or \eqref{SUPP_Uk}. We will treat both examples studied in the main text on an equal footing, where we always refer to the system described in Sec.~\ref{SUPP_Two_atoms} as example (i) and to the setting of Sec.~\ref{SUPP_feedback} as example (ii).

\subsection{Hilbert space in a time bin representation}

The Hilbert space for the full quantum state of atoms and field in the time bin representation, $\ket{\Psi(t)}$, is in both cases $\mathcal{H}=\mathcal{H}_S\bigotimes_{p=-\infty}^\infty \mathcal{H}_p$,
where $\mathcal{H}_S$ is the Hilbert space of the the system [atoms(s)], which in (i) is of dimension $d_S=4$, and in (ii)  it is of dimension $d_S=2$. For future reference we label a basis in the system Hilbert space $\mathcal{H}_S$ by $\ket{i_S}$ ($i_S=1,\dots d_S$). With each time bin $p$ there is an associated Hilbert space $\mathcal{H}_p$, that in (i) contains two bosonic modes corresponding to right and left moving photons, and in (ii) only one [see Fig.\ref{FigSUPP1}(b,d)]. We label a basis in $\mathcal{H}_p$ by $\ket{i_p}$. 
For the example (ii) we can construct such a basis as 
\begin{align}
\ket{i_p}=\frac{(\Delta B^{\dag})^{i_p}}{\sqrt{\Delta t ^{i_p} i_q!}}\ket{{\rm vac}_p},\quad (i_p=0,1,2,\dots)
\end{align} 
and where $\Delta B(t_p)\ket{{\rm vac}_p}=0$. Thus $\ket{i_p}$ denotes the state in which the time bin  $p$ is populated with  $i_p$ photons.
For the example (i) we have 
\begin{align}
\ket{i_p}\equiv \ket{i_p^L,i_p^{R}}=\frac{(\Delta B_L^{\dag})^{i_p^L}}{\sqrt{\Delta t ^{i_p^L} i_q^L!}}\frac{(\Delta B_R^{\dag})^{i_p^R}}{\sqrt{\Delta t ^{i_p^R} i_q^R!}}\ket{{\rm vac}_p},
\end{align}
with ($i_p^{L/R}=0,1,2,\dots$) and where $\Delta B_{L/R}(t_p)\ket{{\rm vac}_p}=0$. Thus, $\ket{i_p}=\ket{i_p^L,i_p^R}$ denotes the state with $i_p^L$ photons in the ``left-moving'' mode of time bin $p$ and $i_p^R$ photons in the ``right-moving'' mode. 
We note that for the system considered here each bosonic mode in this time bin representation interacts with the system (at most) twice for a short time $\Delta t$ in the entire evolution from the initial state at $t=0$ to $t\rightarrow\infty$. Thus, for a waveguide initially in the vacuum state, one can restrict the Hilbert space for each of these modes to a maximum occupation number of two.

\subsection{State of the system and radiation field}

We assume that the state of the system and the radiation field, $\ket{\Psi(t)}\in \mathcal{H}$, is initially ($t=0$) completely uncorrelated, that is 
\begin{align}\label{SUPP_psi0}
\ket{\Psi(t=0)}=\ket{\psi_{S}}\bigotimes_{p=-\infty}^{\infty}\ket{\phi_{p}}
\end{align}
Here $\ket{\psi_{S}}$ denotes the initial state of the system
and $\ket{\phi_{p}}$ the state of the photons in time bin $p$. For concreteness we chose waveguide initially in the vacuum state
$\ket{\phi_{p}}=\ket{{\rm vac}_p}$ (for all $p$).

Under the evolution with the stroboscopic maps $U_k$ [\eqref{SUPP_QSSEmap} or \eqref{SUPP_Uk}] until a time $t_k$ the system interacts sequentially with pairs of time bins $(0,-\ell),(1,1-\ell),(2,2-\ell),\dots (k-1,k-1-\ell)$ as illustrated in Fig.~\ref{FigSUPP1}(b,d). Thus, it gets correlated with the time bins $k-1,k-2,\dots$, but not with the time bins $k,k+1,k+2,\dots$, and the state can be written as $\ket{\Psi(t_{k})}=\ket{\phi_{in}(t_{k})}\otimes\ket{\psi(t_{k})}$,
where $\ket{\phi_{in}(t_{k})}=\bigotimes_{p\geq k}\ket{{\rm vac}_{p}}$ and

\begin{align}\label{SUPP_psit}
\ket{\psi(t_{k})} & =\!\!\!\!\!\sum_{i_{S},\{{i_{p}}\}}\!\!\!\!\psi_{i_{S},i_{k-1},i_{k-2},\dots}\ket{i_{S},i_{k-1},i_{k-2}\dots}
\end{align}
is the entangled state of emitters and radiation field in time bins $p<k$. 

At the time $t_k$ one can identify the state of the time bin $k$ as the input state for the following stroboscopic step. The state of the time bins 
$p\in(k-1,\dots,k-\ell)$ represents the photonic state in the quantum
circuit, while the time bins $p<k-\ell$ correspond to the output field {[}cf.~Fig.~\ref{Fig1}(b){]}.

\subsection{MPS Ansatz}

In a matrix product state representation we can write the coefficients in \eqref{SUPP_psit} as
\begin{align}
\psi_{i_{S},{i}_{k-1},\dots\!}=\tr{A[S]^{i_{S}}A[k-\!1]^{i_{k-1}}A[k-\!2]^{i_{k-2}}\!\!\dots}\label{SUPP_MPS}
\end{align}
or in the canonical form \cite{Schollwock:2011gl}
\begin{align}
&\psi_{i_{S},{i}_{k-1},\dots\!}=\no\\
&=\sum_{\{\alpha\}} \Gamma[S]^{i_{S}}_{\alpha_S}\Lambda[S]_{\alpha_S}\Gamma[k\!-\!1]^{i_{k\!-\!1}}_{\alpha_S,\alpha_{k\!-\!1}}\Lambda[k\!-\!1]_{\alpha_{k\!-\!1}}\cdots \label{SUPP_MPS_canonical}
\end{align}
Here the $\Lambda[S]_{\alpha_S}$ are the Schmidt coefficients for a bipartite splitting of the system from the radiation fields and thus determines the entanglement of the atom(s) with the photons. 
Analogously, $\Lambda[p]_{\alpha_p}$ on the other hand are the Schmidt coefficients for a bipartite splitting of the time bins $q<p$ from the rest of the state consisting of time bins $q\geq p$ and the atom(s).
For example, $\Lambda[k-\ell]_{\alpha_{k-l}}$ is the Schmidt vector for a bipartite splitting on the entire circuit, including the atom(s) and the radiation field in the delay line, from the output field at time $t_k$. We denote the Schmidt rank in such splittings by $D_p$, and the maximum Schmidt rank in the MPS by $D_{\rm max}$.

\begin{figure*}
\centering \includegraphics[width=0.8\linewidth]{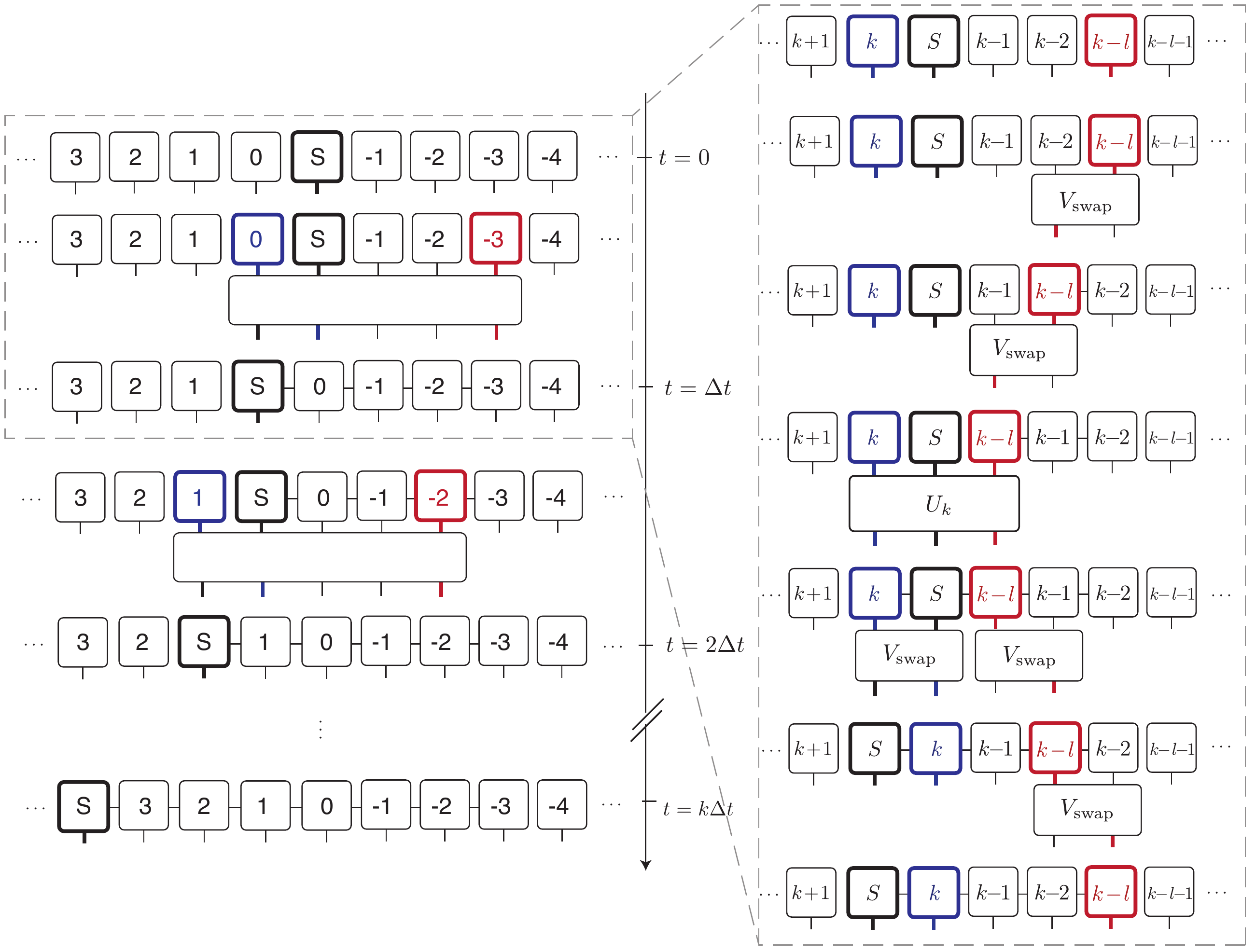} \caption{Illustration for the update of the MPS corresponding to the situations depicted in Fig.~\ref{FigSUPP1}, for $\ell=3$ setting the delay to $\tau=3\Delta t$. Left: At time $t=0$ the full system is in a completely factorized pure state, where the system (S) and the different time bin modes are not entangled (indicated by the absence of vertical links). In the first time step a unitary operation involving the time bin $0$ an the time bin $-\ell$ has to be applied. This is a ``long distance'' operation, which can be implemented by a sequence of local operations as indicated on the right. After the first time step the system is entangled with the time bin $0$ and the time bin $p$ as indicated by the horizontal links. Moreover the system is moved by one site to the left in the MPS representation. By iterating this procedure the system ``moves'' trough the MPS getting entangled with everything that it leaves to its right. Right: MPS update for a single time step. In order to realize the non-local term $U_k$ we apply a sequence of nearest neighbor local unitary operations (from top to bottom). In order to implement the interaction of the system $S$ with the time bin $k$ and the time bin $k-\ell$, that propagates the state $t_k\rightarrow k+1$, $U_k$ we use a sequence of nearest neighbor swap gates that interchange (the quantum state of) time bins, such that $U_k$ can be implemented as a nearest neighbor term. Then the order is restored by the inverse sequence of swap gates. In addition to this we swap the order of the system $S$ with the time bin $k$ such that finally the state after the whole sequence (bottom) is of the same form as before the sequence (top) with $k\rightarrow k+1$.
The entanglement properties indicated by the horizontal bonds correspond to a situation for $k=0$. Note we do not perform at any point a Trotter decomposition of the unitary. }
\label{FigSUPP_MPS} 
\end{figure*}

\subsection{Propagation and MPS update}
As outlined above the interaction of the atoms with the waveguide will in general lead to a highly entangled state. To represent this state efficiently and analyze its entanglement properties we employ MPSs. In the following we outline the method we implement to propagate the MPS according to the stroboscopic map \eqref{SUPP_QSSEmap} or \eqref{SUPP_Uk}. 
\subsubsection{Initial state}
We assume that waveguide is initially ($t=0$) in the vacuum state 
\begin{align}\label{SUPP_psi0_2}
\ket{\Psi(t=0)}=\ket{\psi_{S}}\bigotimes_{p=-\infty}^{\infty}\ket{{\rm vac}_{p}}
\end{align}
which has a trivial MPS representation with bond Dimension $D=1$ (Fig.~\ref{FigSUPP_MPS}). 

\subsubsection{First time step}
In the fist time step $t=0\rightarrow t=\Delta t$ the system interacts with the photonic modes in the bins $p=0$ and $p=-\ell$, such that the system at time $t=\Delta t$ is entangled with these two time bins, while the other modes are still uncorrelated. To update the MPS accordingly one has to implement a ``long range interaction'' between the three involved time bins (Fig.~\ref{FigSUPP_MPS}). To do so we use a method proposed in \cite{Schachenmayer:2010ia}, and reduce it to a sequential application of  nearest neighbor unitary operations, where standard tools can be used to update the MPS accordingly.
As indicated in Fig.~\ref{FigSUPP_MPS} we do this in four steps:

\begin{itemize}
\item
We perform a series of swap gates, $V_{\rm swap}^{(p,p-1)}$, that interchange the quantum state in time bins $p$ and $p-1$. To keep the state in MPS form we have to perform a singular value decomposition after each such nearest neighbor unitary operation. With the application of $\ell-1$ such gates with $p=-\ell-1,-\ell-2,\dots,-1$, we cyclically permute the quantum state in time bins representing the delay line, such that the state of time bin $-\ell$ is moved ``next to the system'' (see Fig.~\ref{FigSUPP_MPS}).  
\item
After this permutation the unitary $U_{k=0}$ can be applied in a local way to the state of system, time bin $p=0$ and time bin $p=-\ell$.
\item
We reorder the state by swapping back the time bin $-\ell$ to its original position. 
\item
To prepare the state for the next time step ($k=1$) we bring the state to the form \eqref{SUPP_MPS_canonical} (with $k=1$) by a swap of the system with the time bin $0$. 
\end{itemize}
In summary, this involves $2\ell-1$ local unitary operations (singular value decompositions) to propagate the state for a single time step and update the tensors $\Gamma[S]$, and $\Gamma[p]$ ($p=-1,-2,\dots -\ell$)and the Schmidt vectors $\Lambda[S]$ and $\Lambda[p]$  ($p=-1,-2,\dots -\ell+1$). The entanglement between the system ant the two time bins $0$ and $-\ell$ is reflected by an increase of the bond dimension $D_S$ and $D_p$ for $p=-1,-2,\dots -\ell+1$ (indicated by the horizontal links in Fig.~\ref{FigSUPP_MPS}).

\subsubsection{k-th timestep}

The propagation of the next steps $t_k\rightarrow t_{k+1}$ can be achieved in the same way by applying:
\begin{itemize}
\item
A series of local swap gates, $V_k=\prod_{p=k-1}^{k-\ell+1} V_{\rm swap}^{(p,p-1)}$.  
\item
The unitary $U_{k}$  (local).
\item
The series of local swaps $V_k^{\dag}$ to restore the order of the MPS.
\item
The local swap gate $V_{\rm swap}^{(k,S)}$ to prepare the MPS for the next time step.
\end{itemize}
Again $2\ell-1$ singular value decompositions have to be performed to update the $\ell$ tensors $\Gamma[S]$, and $\Gamma[p]$ and the Schmidt vectors $\Lambda[S]$ and $\Lambda[p]$ for $p=k-1,k-2,\dots k-\ell$ in each step. We note in particular that the tensors $\Gamma[p]$ and the Schmidt vectors $\Lambda[p]$ for $p<k-l$, that is for time bins representing the output field are unchanged in the $k$-th time step (and in all steps after that), that is, once a time bin leaves the circuit its state  does not change anymore. 

\begin{figure}[h]
\centering \includegraphics[width=0.5\textwidth]{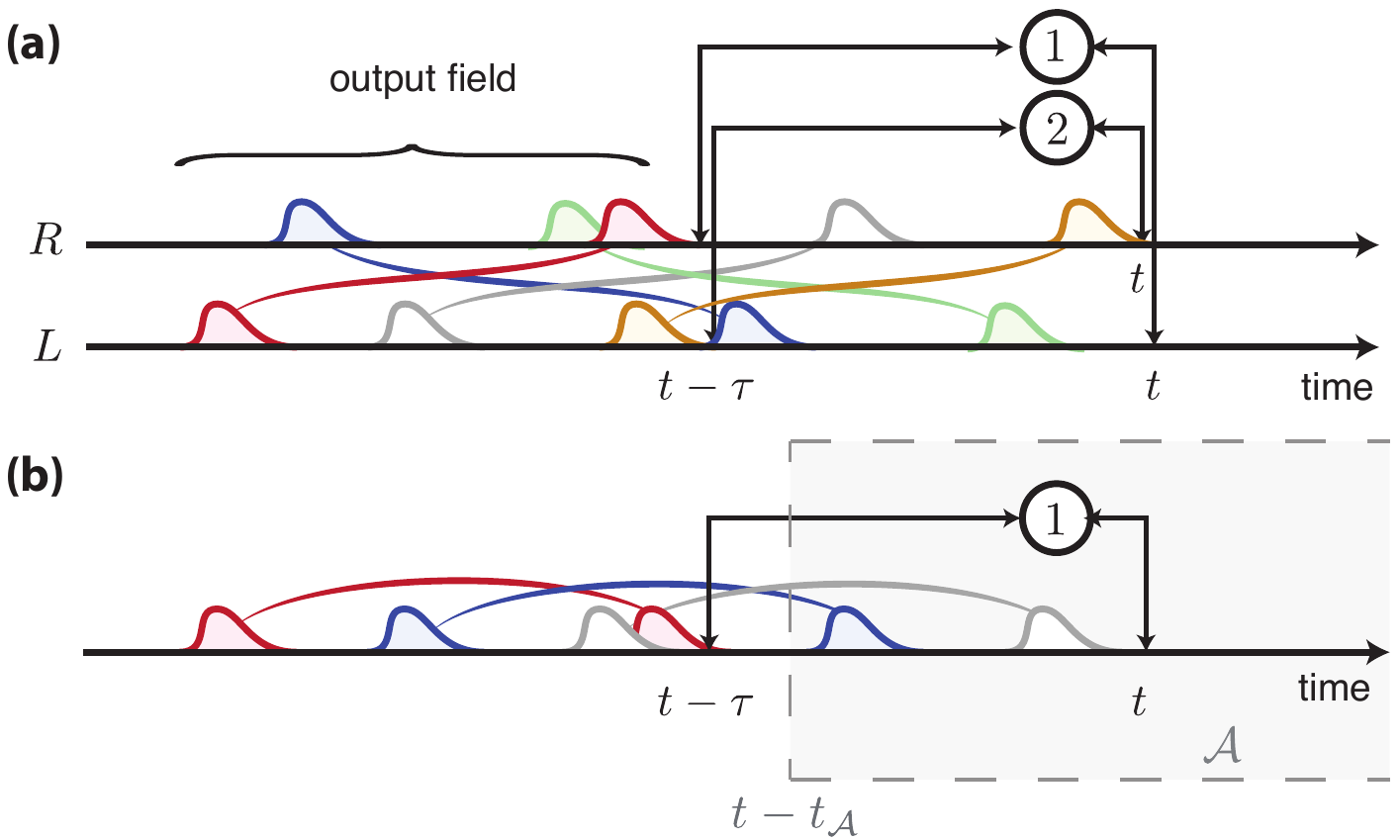} \caption{Schematic representation of the entanglement generated for two atoms coupled to a bidirectional waveguide (a) and a single atom to a mirror-ended waveguide (b). The photons emitted each are in an entangled state between left moving lime time bins and right moving time bins. The time delay causes entanglement of this photon in the time basis over a time $\tau$. For a bipartition of the system at a time bin $t-t_\mathcal{A}$ the entanglement entropy of the field and the system in $\mathcal{A}$ with the rest of the field is proportional to the links cut in this splitting (e.g. blue and grey in this example). The maximum entanglement entropy for such a splitting is thus proportional to the number of photons emitted over a time $\tau$. The maximum entropy in the output field obeys thus an area law.}
\label{FigSUPP_Entropy} 
\end{figure}

As discussed in the main text, the necessary bond dimension $D_{\rm max}$ to represent the state of the radiation field in this MPS form depends only on the time delay $\tau$ and scales as $D_{\rm max}\sim 2^{\gamma \tau S_1}$, where $S_1$ is the entropy per photon defined in the main text. This sets the limit on the achievable delay time $\tau$ for a given bond dimension $D_{\rm max}$. On the other hand, for a fixed $\tau$ the bond dimension does not increase with the total integration time $T_{\rm max}=k_{\rm max}\Delta t\gg \tau$ (See also Fig.~\ref{FigSUPP_Entropy}). This allows us access the steady state by propagating the full state to a time $T_{\rm max}$, where the properties of the reduced state of the circuits relax to their stationary value.

\subsection{Observables and state properties}

With the full quantum state in the canonical matrix product form \eqref{SUPP_MPS_canonical} we have access to various quantities of both, the system and the radiation field. In particular this includes the entanglement of various bi-partitions of the full system, e.g. in terms of the entanglement entropy, but also properties of the output field that are accessible in experiments. We are both interested in the time evolution of these quantities, as well as in their values in the steady state. We access latter by calculating the full time evolution up to a (large) time $T_{\rm max}=k_{\rm max}\Delta t$ for which they relaxed to their stationary value. 

\textit{Entanglement Entropy:}
In the main text we quantify the entanglement at a time $t_k$ for a bipartition of the entire system in a part $\mathcal{A}$ that contains the system Hilbert space and the radiation fields in the interval $[t_k,t_k-t_\mathcal{A}]$ with the rest of the system, that is
\begin{align}
\mathcal{A}=\mathcal{H}_S\!\!\!\!\bigotimes_{k>p\geq k-p_{\mathcal{A}}}\!\!\!\!\mathcal{H}_p
\end{align}
with $t_\mathcal{A}\equiv p_\mathcal{A}\Delta t$. From Eq.~\eqref{SUPP_MPS_canonical} the spectrum of the corresponding reduced Hilbert space is given by $\Lambda[k-p_\mathcal{A}]^2 $. The corresponding entanglement entropy is then given by 
\begin{align}
S(\rho_{\mathcal{A}})=-\sum_{\alpha}\Lambda[k-p_\mathcal{A}]_{\alpha}^2\log_2( \Lambda[k-p_\mathcal{A}]^2_{\alpha})
\end{align}
Of particular interest are the entanglement of the atoms with the radiation field, that is the entropy of the reduced state of the atoms $\rho_{\rm sys}$:
\begin{align}
S(\rho_{\rm sys})=-\sum_{\alpha}\Lambda[S]_{\alpha}^2\log_2( \Lambda[S]^2_{\alpha})
\end{align}
but even more the entanglement of the entire photonic circuit (including atoms and radiation fields in the delaylines) with the output field 
\begin{align}
S(\rho_{\rm circuit})=-\sum_{\alpha}\Lambda[k-\ell]_{\alpha}^2\log_2( \Lambda[k-\ell]^2_{\alpha}).
\end{align} 
We note that the entanglement does not depend on the choice of the time step.

\textit{Spectrum of the output field:} In the steady state the spectrum of the output field is given by
\begin{align}
\mathcal{S}_{i,j}(\nu)&=2\Re \int_0^\infty dt' \mean{b_i(t)^\dag b_j(t-t')}e^{i\nu t'}
\end{align}
in the time bin basis this is replaced by
\begin{align}\label{SUPP_Spec}
\mathcal{S}_{i,j}(\nu)&\rightarrow2\Re\frac{1}{\Delta t}\sum_{p=0}^{M-1}\mean{\Delta B_i^\dag(t_q)\Delta B_j(t_{q-p})}e^{i\nu p \Delta t}
\end{align}
where $M$ has to be chosen large enough and $q=k_{\rm max}-\ell-1$, such that the correlation function in \eqref{SUPP_Spec} is evaluated for the output field at the time $t_k$. The correlation function appearing in \eqref{SUPP_Spec} can be efficiently evaluated for a state in the form \eqref{SUPP_MPS_canonical}.

\textit{Higher order photon correlation functions:} In the same manner as for the spectrum one can also evaluate higher order correlation functions of the output field that are experimentally accessible such as the second order autocorrelation function, which in steady state ($t\rightarrow \infty$) is defined as
\begin{align}
g_2(t')=\frac{\mean{b(t)^\dag b^\dag(t-t')b(t-t')b(t)}}{\mean{b^\dag(t)b(t)}^2}
\end{align}
which again in the time bin basis translates to 
\begin{align}
g_2(p\Delta t)=\frac{\mean{\Delta B(t_q)^\dag \Delta B^\dag(t_{q-p})\Delta B^\dag(t_{q-p})\Delta B(t)}}{\mean{\Delta B^\dag(t_q)\Delta B(t_q)}^2}
\end{align}
where $q=k_{\rm max}-\ell-1$.

\subsection{Generalizations and Extensions}

\subsubsection{Multiple Nodes and channels}

The above discussion can be generalized to more photonic channels in a straightforward way, essentially extending the number of modes in each time bin, c.f.~Fig.~\ref{Fig1}(b). Each channel adds another leg to the ``conveyor belt'' carrying the photonic excitations. Additional 
nodes can be added by noting that for independent nodes the stroboscopic map can be written as a
product of maps for the individual nodes $U_{k}=\prod_{n}U_{k,n}$. At
this point it is also easy to include various optical elements in the waveguides,
e.g.~beam splitters, by terms of the form $U_{k,n}=\exp(-i(\chi\Delta B_{a_{m}}^{\dag}(t_{c_{m}})\Delta B_{b_{m}}(t_{d_{m}})-{\rm h.c.))}$.

\subsubsection{Nonlinear dispersion relation}

We note that the treatment presented here is tailored to describe
transmission lines with a {\em linear} dispersion relation. We can however relax this assumption and
generalize this to include deviations thereof. A quadratic contribution,
that is $\omega(k)\approx vk+\alpha(k-\bar\omega/v)^{2}$ can
be modeled by evolving the radiation field in each time step in addition to $U_k$ also with an additional Hamiltonian term of the form $$H_{\rm dis}\!=-\frac{\hbar J}{\Delta t}\sum_{p}(\Delta B^{\dag}(t_p)\Delta B(t_{p+1})-\Delta B^{\dag}(t_p)\Delta B(t_p)+{\rm {h.c.})},$$
with $J=\alpha/(v\Delta t)^{2}\ll1/\Delta t$.
This is a nearest neighbor term and can be included in the
MPS evolution in a standard way. 

\subsubsection{System reservoir coupling}
In the derivation of \eqref{Eq2-1} we approximated the couplings of the atomic dipole to a photon of frequency $\omega$, $\kappa_i(\omega)$, by a constant. One can again go beyond this and model a situation with $\kappa_i(\omega)=\kappa_i(\bar \omega)+\beta_i(\omega-\bar \omega)$. In the example of Sec. \ref{SUPP_feedback} this would lead to an additional term ${H}_{{\rm int}, I}(t)\rightarrow {H}_{{\rm int}, I}(t)+{H}_{{\rm int}, I}^{(1)}(t)$ with
 \begin{align}
{H}_{{\rm int}, I}^{(1)}(t)=&i\hbar\lr{\beta_R i\frac{d}{dt}b(t)+\beta_L i\frac{d}{dt}b^{\dag}\lr{t-\tau}e^{i\phi})c-\textrm{h.c.}}.
\end{align}
In terms of the stroboscopic map \eqref{SUPP_Uk} this would correspond to an additional term $O_k\rightarrow O_k+O_k^{(1)}$ with
 \begin{align}
O_k^{(1)}=&\frac{i\beta_R}{\Delta t} (\Delta B^\dag(t_k)-\Delta B^\dag(t_{k-1}))c\no\\
&+\frac{i\beta_L}{\Delta t}(\Delta B^\dag(t_{k-\ell})-\Delta B^\dag(t_{k-\ell-1}))e^{i\phi})c-\textrm{h.c.}.
\end{align}
Again this could be straightforwardly included in the MPS description.

\begin{figure}[h]
\centering \includegraphics[width=0.5\textwidth]{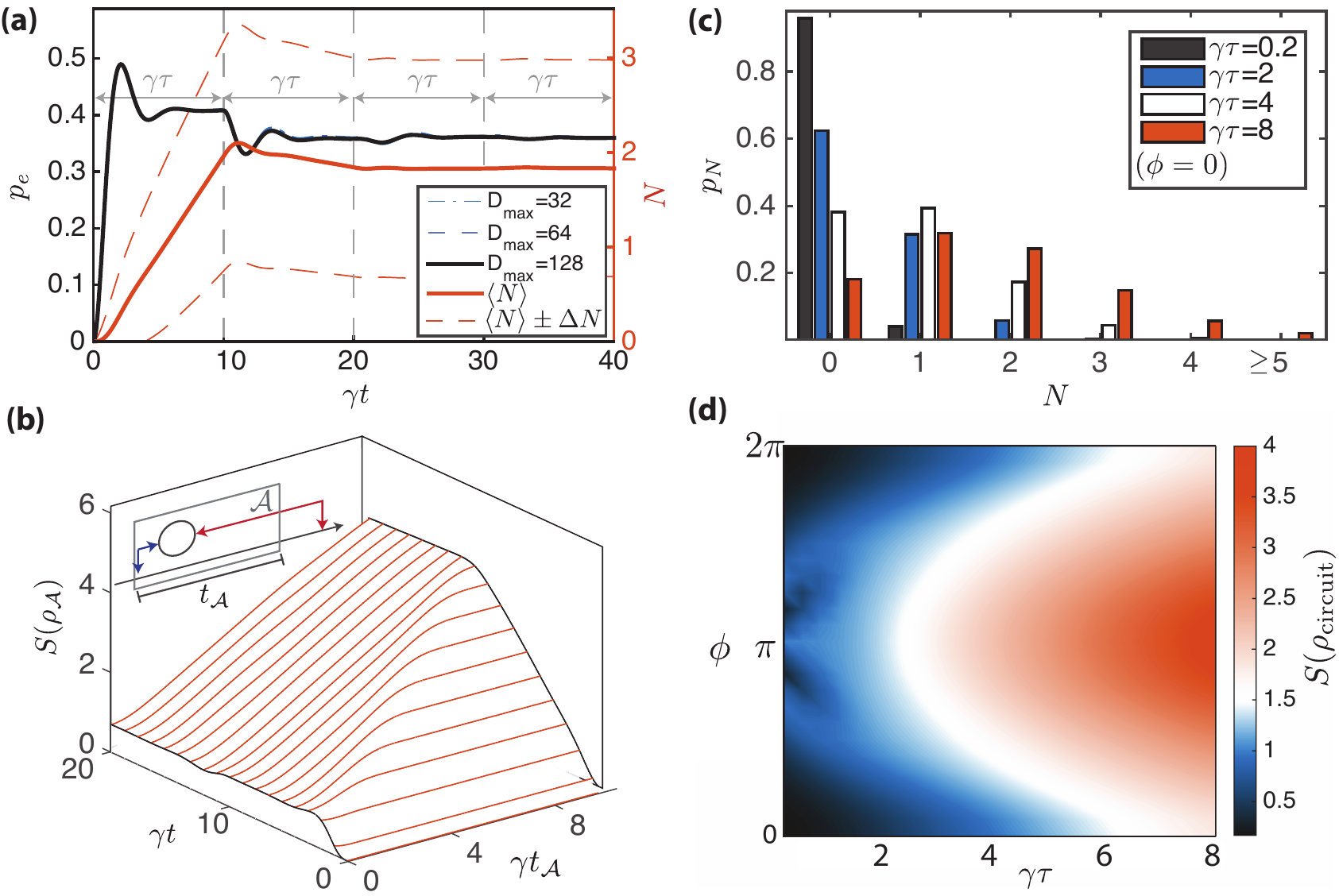} \caption{Driven atom in  front of a distant mirror (compare to Fig~\ref{Fig2} of the main text and Fig.~\ref{FigSUPP1}) . (a) Time evolution of the atomic
excitation probability (black lines) and photon number in the waveguide between the atom and the mirror (red). Multiples of the delay time $\tau=10/\gamma$ are indicated by vertical dashed lines. Parameters:$\Omega/\gamma=1.5$, $\gamma_L=\gamma_R\equiv\gamma/2$, $\phi=0$, $\Omega=1.5\gamma$, $\Delta=0$; $\gamma_R\Delta t=0.05$.
We plot $p_{e}$ for different values of for the maximal bond dimension $D_{\rm max}=32,64,128$ of the MPS to illustrate convergence. 
(b) Entanglement entropy of the system and radiation field in in the interval $[t,t-t_\mathcal{A}]$ as a function of time. The line at $t_{\mathcal{A}}=0$ gives the entanglement
of the atom with the entire radiation field, while the line at $t_{\mathcal{A}}=\tau=10/\gamma$
corresponds to the entanglement of the entire circuit with the output field. 
Parameters are the same as in (a).
(c) Probabilities $p_{N}$ for
having $N$ photons in the waveguide between the atom and the mirror for different delay times
$\tau$ and a coherent drive $\Omega/\gamma=2.5$ in the steady state (calculated by time evolution to $t_{\rm max}=200/\gamma$). 
(d) Entanglement entropy of the entire circuit with the output field in the steady
state. Parameters as in Fig.~\ref{Fig4} of the main text. }
\label{FigSUPP_Time} 
\end{figure}

\section{Additional Results: Setup 2}

To complement the results shown in the main text we report here results on the quantum feedback problem, where we plot analogous quantities to the ones calculated in Fig.~\ref{Fig2} of the main text for the problem of delayed interaction of two atoms via a waveguide. 

In Fig.~\ref{FigSUPP_Time}(a) we consider an atom initially in the ground state and the waveguide in the vacuum. The mirror is place at a distance $d$ such that $\gamma \tau=10$ and $\phi=0$. We show the time evolution of the atomic excitation probability $p_e=\textrm{Tr}\{\ket{e}\bra{e}\ket{\Psi(t)}\bra{\Psi(t)}\}$ for a system that is driven coherently with $\Omega/\gamma=3/2$ and on  resonance $\Delta=0$. For times $t<\tau$ the atom does not ``see the mirror'' and evolves according to the standard optical Bloch equations that govern the dynamics of the reduced atomic state in the absence of the mirror. For times $t>\tau$ the delay non-classical steam of photons that bounced off the mirror ads to the coherent driving filed. With the choice of $\phi=0$ this feedback is in phase with the coherent drive and enhances emission into the output port. Due to this enhanced emission the excitation probability of the atom is reduced after the first roundtrip. In Fig.~\ref{FigSUPP_Time}(a) we also plot the mean number of photons between the half cavity formed by the atom and the mirror, and fluctuations. 

Fig.~\ref{FigSUPP_Time}(b) shows how the entanglement builds up during the time evolution. During the first round trip the $S(\rho_{\rm circuit})$ increases approximately linear with time reflecting the linear increase of the number of photons that are emitted in an entangled  state towards the mirror and away from it. After the first roundtrip the entanglement entropy reaches essentially a steady state. The linear increase stops, since for each entangled photon that is created during the second round trip time, on average one photon that was emitted during the first round trip time leaves the half cavity and does no longer contribute to the entanglement entropy (see also Fig.~\ref{FigSUPP_Entropy}(b)]. 

Fig.~\ref{FigSUPP_Time}(c) shows the number statistics of photons in the feedback loop in steady state. Clearly for increasing $\tau$ the number increases showing that our approach allows us to go clearly beyond the Markovian approximation. 

Finally, in Fig.~\ref{FigSUPP_Time}(d) we show the entanglement entropy $S(\rho_{\rm circuit})$ in the steady state. As argued above, the entanglement increases approximately linear with the length of the delay line. The dependence on $\phi$ is related to the excitation probability of the atom in the steady state [see Fig.~\ref{Fig4}(a)]. For $\gamma\tau\gtrsim 1$ it governs the emission rate of entangled photons towards and away from the mirror. For $\gamma\tau
\lesssim 1$ the feedback field interferes destructively with the coherent drive if the feedback phase conspires with the delay time to  $\phi=\pi-\sqrt{\Omega^{2}+\Delta^{2}}\tau$. In this case that the atom  reabsorbs the photons reflected from the mirror, and no photons are emitted into the output port in steady state [see Fig.\ref{Fig4}(a)], explaining the correspondingly low entanglement in Fig.~\ref{FigSUPP_Time}(d).
 

\end{document}